\documentclass[prx,aps,twocolumn,preprintnumbers,superscriptaddress,showpacs]{revtex4-1}
\usepackage{bbm}
\usepackage{mathrsfs}
\usepackage{epsfig}
\usepackage{graphicx}
\usepackage{amsfonts}
\usepackage[figuresright]{rotating}
\usepackage{amssymb}
\usepackage{amsmath}
\usepackage{dcolumn}
\usepackage{bm}
\usepackage{color}
\usepackage{psfrag}
\usepackage{natbib}
\usepackage{graphicx,amsfonts,amssymb,amsmath}
\usepackage{overpic}

\begin{document}

\title{A generalized Lanczos method for systematic optimization of tensor network states}

\author{Rui-Zhen Huang}
\affiliation{Institute of Physics, Chinese Academy of Sciences, P.O. Box 603, Beijing
100190, China}

\author{Hai-Jun Liao}
\affiliation{Institute of Physics, Chinese Academy of Sciences, P.O. Box 603, Beijing
100190, China}

\author{Zhi-Yuan Liu}
\affiliation{Institute of Theoretical Physics, Chinese Academy of Sciences, P.O. Box 2735, Beijing 100190, China}

\author{Hai-Dong Xie}
\affiliation{Institute of Physics, Chinese Academy of Sciences, P.O. Box 603, Beijing
100190, China}

\author{Zhi-Yuan Xie}
\affiliation{Department of Physics, Renmin University of China, Beijing 100872, China }

\author{Hui-Hai Zhao}
\affiliation{Department of Applied Physics, University of Tokyo, Hongo, Bunkyo-ku, Tokyo 113-8656, Japan}
\affiliation{Institute for Solid State Physics, University of Tokyo, Kashiwanoha, Kashiwa, Chiba 277-8581, Japan}

\author{Jing Chen}
\affiliation{Institute of Physics, Chinese Academy of Sciences, P.O. Box 603, Beijing 100190, China}

\author{Tao Xiang}
\email{txiang@iphy.ac.cn}
\affiliation{Institute of Physics, Chinese Academy of Sciences, P.O. Box 603, Beijing
100190, China}
\affiliation{Collaborative Innovation Center of Quantum
Matter, Beijing 100190, China}

\begin{abstract}
  We propose a generalized Lanczos method to generate the many-body basis states of quantum lattice models using tensor-network states (TNS).
  The ground-state wave function is represented as a linear superposition composed from a set of TNS generated by Lanczos iteration. This method improves significantly both the accuracy and the efficiency of the tensor-network algorithm and allows the ground state to be determined accurately using TNS with very small virtual bond dimensions. This state contains significantly more entanglement than each individual TNS, reproducing correctly the logarithmic size dependence of the entanglement entropy in a critical system. The method can be generalized to non-Hamiltonian systems and to the calculation of low-lying excited states, dynamical correlation functions, and other physical properties of strongly correlated systems.
\end{abstract}

\maketitle

One of the biggest challenges and unsolved problems in condensed matter and quantum field theory is the study of quantum many-body systems, whose state space is exponentially large. This has severely delayed our understanding of many fascinating strongly correlated phenomena, including high temperature superconductivity and quantum spin liquids.  Quantum Monte Carlo is one of the most successful methods in simulating quantum many-body systems, but fails in the simulation of interacting fermion and frustrated spin models due to the infamous minus sign problem. In recent years, tremendous progress has been achieved in the development of numerical renormalization-group methods based on TNS \cite{1997NiggemannKlumperZittartz,2000Nishino,2004VerstraetePEPS,2007LevinNave,2007VidalMERA,
2008JiangSimpleUpdate,2008Jordan,2009XieSRG,2012XieHOTRG,2014Corboz-tJmodel,2014XiePESS}, which have emerged as a powerful theoretical tool for investigating low-dimensional quantum lattice models.

The TNS formulation is a variational ansatz for the ground state. It reduces the dimension of the Hilbert space, which grows exponentially with system size, to a polynomial grouth. Commonly used TNS include the one-dimensional matrix-product state (MPS) \cite{1995Ostlund-PRL}, which is a class of states underlying the density-matrix renormalization group \cite{1992WhiteDMRG}, and the two-dimensional projected entangled pair state (PEPS) \cite{2004VerstraetePEPS}. The accuracy of TNS is controlled by the virtual bond dimension of the local tensors, $D$. The larger is the bond dimension, the more accurate is a TNS. However, the cost for computing a TNS, especially a PEPS or a projected entangled simplex state (PESS) \cite{2014XiePESS}, rises rapidly with $D$. For example, the minimal cost scales as $D^{12}$ for PEPS. This has limited the bond dimension that can be handled to be generally less than 13 in two dimensions \cite{2008JiangSimpleUpdate,2014XiePESS,2014Corboz-tJmodel}. Furthermore, although both MPS and PEPS satisfy the area law of entanglement entropy \cite{2010EisertAreaLaw}, in a critical or interacting fermion system with a finite Fermi surface, there is a logarithmic correction to this entropy. To describe correctly this logarithmic behavior, a more complex TNS structure is requred, namely the multi-scale entanglement renormalization ansatz (MERA)  in one dimension \cite{2007VidalMERA} or the branching MERA in two dimensions \cite{2014EvenblyBranchMERA}. The cost for handling these MERA-type wavefunctions is even higher. Resolving this difficulty requires a new approach that can improve significantly the accuracy of TNS without relying on the increase of bond dimension.

In this letter, we propose a generalized Lanczos method to solve quantum lattice models using TNS. This method is an adaptation of the Lanczos method for the tensor network algorithm, which generates a set of orthonormal many-body basis states (i.e. the Krylov subspace), represented using TNS, by applying the Hamiltonian to the iteratively generated basis states. At each iteration step, a new TNS is generated by minimizing a cost function. However, as a TNS is only an approximate representation of a quantum state, the Hamiltonian is not tri-diagonalized. By diagonalizing the Hamiltonian in this set of basis states, the ground state can be accurately represented as a linear superposition of all the generated TNS. Using the Heisenberg model on a chain and a square, we demonstrate that the results obtained with this method are extremely accurate, even with very small $D$.

{\it Method:}
The TNS-Lanczos method starts from a TNS, $\left\vert \psi_{1} \right\rangle $, that is determined by variationally minimizing the energy functional
\begin{equation}
f\left( \psi _{1}\right) =\left\langle \psi _{1}\left\vert H\right\vert \psi
_{1}\right\rangle ,
\end{equation}
subject to the constraint $\left\langle \psi _{1}|\psi _{1}\right\rangle =1$.
$\left\vert \psi _{1}\right\rangle $ is set as the first normalized many-body basis state, $\left\vert \Psi _{1}\right\rangle =\left\vert \psi _{1}\right\rangle $.
A new TNS, $\left\vert \psi _{2}\right\rangle $, with the
same bond dimension is then generated by minimizing the cost function
\begin{equation}
g\left( \psi _{2}\right) =\left\vert \left\vert H\left\vert \Psi
_{1}\right\rangle -h_{11}\left\vert \Psi _{1}\right\rangle -\left\vert \psi
_{2}\right\rangle \right\vert \right\vert ^{2},  \label{Eq:CostFunc1}
\end{equation}
where $h_{11}$ is the energy of the basis state $\left\vert \Psi_{1}\right \rangle $, defined by Eq. (\ref{Eq:H-matrix-elements}). From $\left\vert \psi _{2}\right\rangle $, we can construct a new normalized basis
state $\left\vert \Psi _{2}\right\rangle $ that is orthogonal to $\left\vert
\Psi _{1}\right\rangle $,
\begin{equation}
\left\vert \Psi _{2}\right\rangle =a_{2}\left( 1-\left\vert \Psi
_{1}\right\rangle \left\langle \Psi _{1}\right\vert \right) \left\vert \psi
_{2}\right\rangle ,
\end{equation}
where $a_{2}$ is the normalization constant.

Similarly, from $\left\vert \Psi _{1}\right\rangle $ and $\left\vert \Psi
_{2}\right\rangle $, we can generate another TNS, $
\left\vert \psi _{3}\right\rangle $, by minimizing a cost function similar to
Eq. (\ref{Eq:CostFunc1}).  Again, from $\left\vert \psi _{3}\right\rangle $,
we can construct a new normalized basis state $\left\vert \Psi
_{3}\right\rangle $ that is orthogonal to both $\left\vert \Psi
_{1}\right\rangle $ and $\left\vert \Psi _{2}\right\rangle $. By continuing
this iteration, we create a set of orthonormal basis states $\left\{
\left\vert \Psi _{\alpha }\right\rangle ;\alpha =1,\cdots k\right\} $ with $
\left\langle \Psi _{\alpha }|\Psi _{\beta }\right\rangle =\delta _{\alpha
,\beta }$.

In general, to find the basis state $\left\vert \Psi _{\alpha
+1}\right\rangle $, we first generate a TNS, $\left\vert \psi _{\alpha +1}\right\rangle $, by minimizing the cost function
\begin{equation}
g\left( \psi _{\alpha +1}\right) =\left\vert \left\vert  H\left\vert \Psi
_{\alpha }\right\rangle -\sum_{\beta \le \alpha } h_{\alpha \beta
}\left\vert \Psi _{\beta }\right\rangle -\left\vert \psi _{\alpha
+1}\right\rangle \right\vert\right\vert^{2},  \label{Eq:CostFunc2}
\end{equation}
where $h_{\alpha \beta }$ is the matrix element of the Hamiltonian in this
set of basis states,
\begin{equation}
h_{\alpha \beta }=\left\langle \Psi _{\alpha }\left\vert H\right\vert \Psi
_{\beta }\right\rangle .  \label{Eq:H-matrix-elements}
\end{equation}
$\left\vert \Psi _{\alpha +1}\right\rangle $ is then defined by
\begin{equation}
\left\vert \Psi _{\alpha +1}\right\rangle =a_{\alpha +1}\left(
1-\sum_{\beta \le \alpha}\left\vert \Psi _{\beta}\right\rangle \left\langle \Psi_{\beta}\right\vert \right) \left\vert \psi _{\alpha +1}\right\rangle ,
\end{equation}
with $a_{\alpha +1}$ the normalization constant.

\begin{figure}[tbp]
\includegraphics[angle=0,scale=0.5]{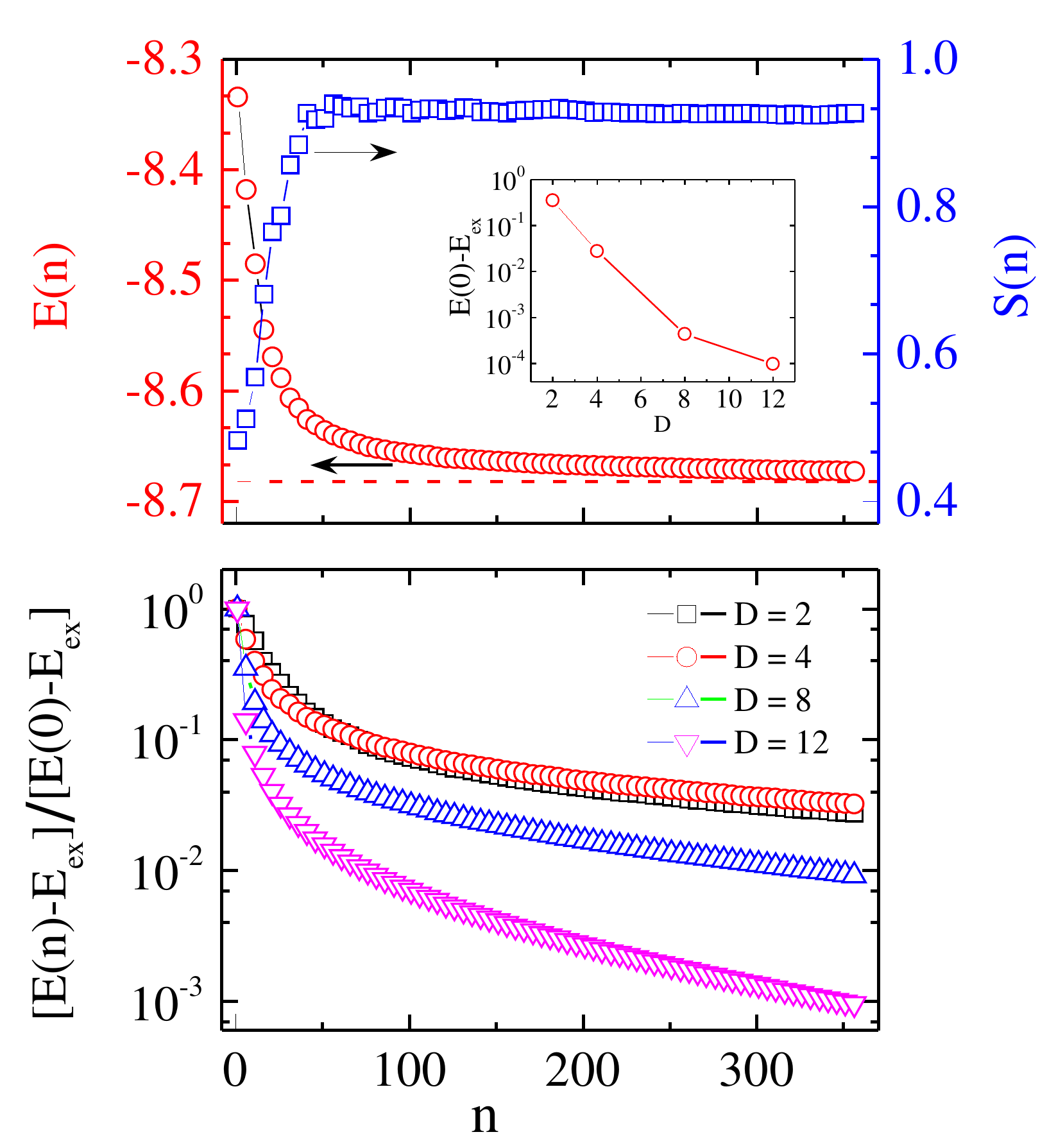}
  \caption{Ground-state energy, $E(n)$, and entanglement entropy, $S(n)$, for the Heisenberg spin chain with $L=20$.  Upper panel: $E(n)$ and $S(n)$ versus $n$ for $D=2$. The red dashed line is the exact ground-state energy, $E_{ex}=-8.6824733344$. The inset shows the difference between the ground state energy obtained with only one MPS, $E(0)$, and $E_{ex}$, as a function of $D$. Lower panel: relative change of ground-state energy, $\left[ E(n)-E_{ex}\right] / \left[ E(0) -E_{ex} \right]$.
  }
\label{fig:Energy1D}
\end{figure}

From the above iteration, we find $k$ TNS, $\left\{
\left\vert \psi _{1}\right\rangle ,\left\vert \psi _{2}\right\rangle ,\cdots
\left\vert \psi _{k}\right\rangle \right\} $ and $k$ orthonormal basis
states $\left\{ \left\vert \Psi _{1}\right\rangle ,\left\vert \Psi
_{2}\right\rangle ,\cdots \left\vert \Psi _{k}\right\rangle \right\} $ in one round of Lanczos iterations. In this basis space, the Hamiltonian can be represented as a $k\times k$ matrix whose matrix elements are defined by Eq. (\ref{Eq:H-matrix-elements}). By diagonalizing this matrix, we obtain the ground-state energy and the corresponding eigenfunction. However, the computational time scales as $\left( k-1\right)
^{2}$, and furthermore, due to the
accumulation of machine errors, the orthogonality of the basis states so
generated may be lost when $k$ becomes large. To avoid these problems, it is better not to use a large $k$ in the calculation, and instead, we use a relatively small $k$ but repeat the above steps many times. Each time we set the ground state obtained from the previous cycle as the initial basis state, $\left\vert \Psi _{1}\right\rangle $. Of course, starting from the second round of iterations, the initial basis state is no longer a pure TNS. Instead it is a linear superposition of all the TNS obtained previously. This restarted Lanczos iteration can be repeated many times, until the ground-state energy converges. If we use $n$ to denote the number of restarted Lanczos iterations, then the total number of TNS generated in this way equals $N=n(k-1)+1$. The ground-state wave function is a linear superposition of these $N$ TNS, $\left\{ |\psi_\alpha\rangle , \, \alpha = 1 \cdots N \right\}$, which we call a Lanczos-generated TNS (abbreviated as LTNS).

A LTNS can be generally expressed as
\begin{equation}
\left\vert \Psi \right\rangle =\sum_{\alpha =1}^{N}c_{\alpha }\left\vert
\psi _{\alpha }\right\rangle ,
\end{equation}
where $c_{\alpha }$ is the coefficient of the ground-state wave function in this tensor-network representation. This wave function can also be represented as a single TNS.
If $A_{\alpha }^{i}\left[ m_{i}\right] $ is the local tensor of $ c_\alpha \left\vert \psi_{\alpha }\right\rangle $, with $m_{i}$ the
quantum number of the physical basis state at site $i$, then the local tensor of
$\left\vert \Psi \right\rangle $ at the same site $i$, $A^{i}\left[ m_{i}
\right] $, is simply a block-diagonal tensor defined by
\begin{equation}
A^{i}\left[ m_{i}\right] =\mathrm{dia}(A_{1}^{i},A_{2}^{i},\cdots ,A_{N}^{i})
\left[ m_{i}\right] .
\end{equation}
The bond dimension of $A^{i}$ equals $ND$. Clearly, $|\Psi \rangle $
contains more variational parameters than each pure TNS, $
\left\vert \psi _{\alpha }\right\rangle $. Thus it is not surprising that it can
be more accurate than the wave function, $\left\vert \psi
_{1}\right\rangle $, obtained by simply minimizing the ground-state energy.

The memory space needed for storing a LTNS scales linearly with $N$. The computational time required for generating these basis states scales as $n^{2}(k-1)^{2} = (N-1)^2$. The converged ground-state energy in the large-$N$ limit depends on the total number of TNS generated, but does not depend much on the value of $k$ also. In the case that only the ground state is studied, it is sufficient to take $k=2$. For a larger $k$, the ground state energy can converge faster than the $k=2$ case during the first tens of iteration, but the entire cost is higher. The results presented below are all obtained with $k=2$.

\begin{figure}[tbp]
\includegraphics[angle=0,scale=0.28]{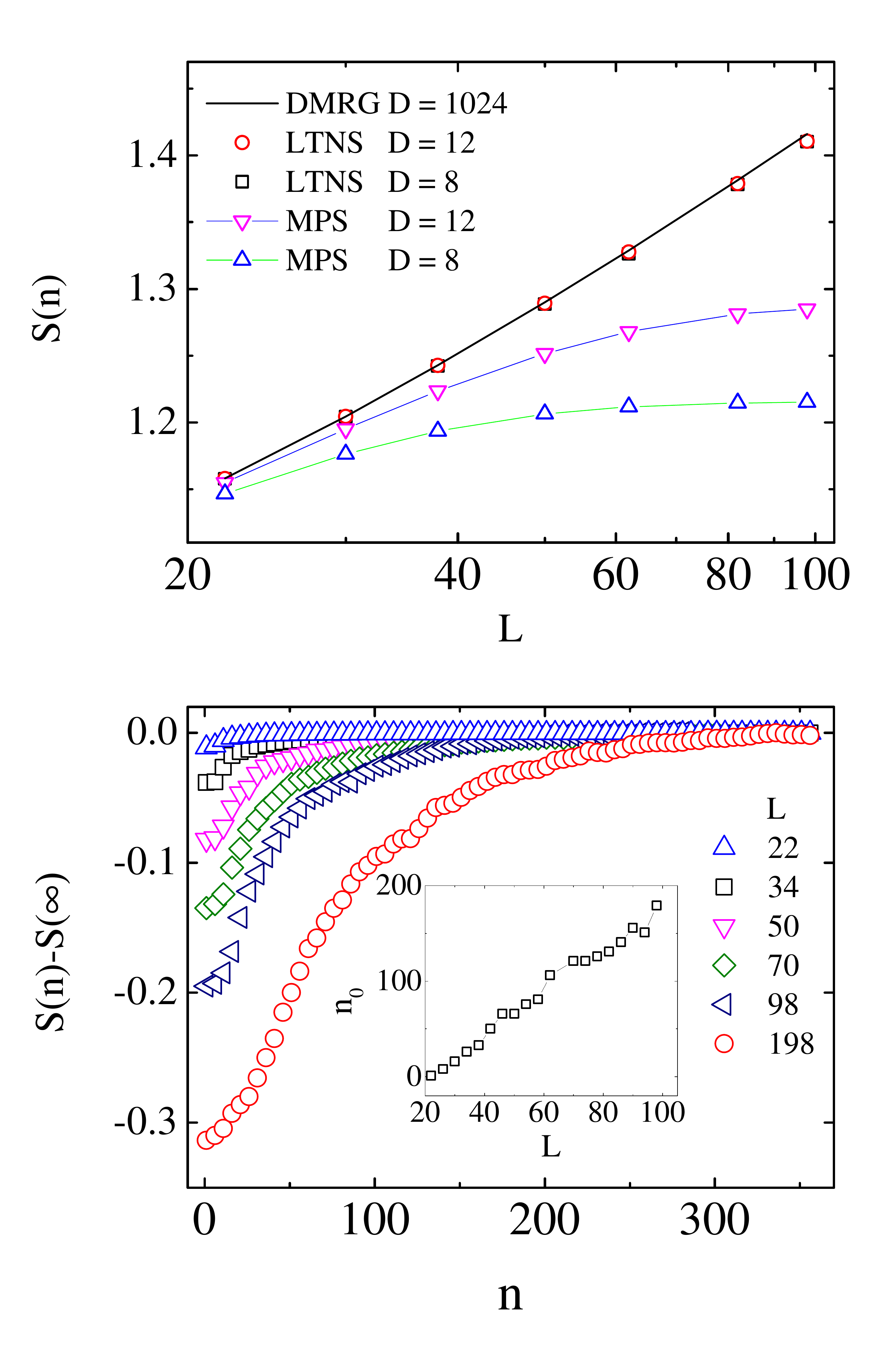}
  \caption{Entanglement entropy for the ground state of the Heisenberg spin chain. (a) Converged entanglement entropy, $S(\infty ) = S(n\rightarrow \infty)$, as a function of $L$ with $D=8$ (squares) and 12 (circles), compared with the corresponding results obtained using a single MPS, $S(n=0)$ (up and down triangles), and those obtained with DMRG by keeping 1024 states (solid line). The horizontal axis is logarithmic.
  (b) Difference between $S(n)$ and $S(\infty )$ as a function of $n$, which shows how fast $S(n)$ converges, for $D=8$. The inset shows the size-dependence of the restarted Lanczos number, $n_0$, at which the entanglement entropy arrives within 1\% of its converged value, i.e. $1 - S(n_0)/S(\infty) \le 1\%$.
  }
\label{fig:Entropy1D}
\end{figure}

{\it Results: }
We have first tested the TNS-Lanczos method using the $S=1/2$ antiferromagnetic Heisenberg spin chain with open boundary conditions.
Fig. \ref{fig:Energy1D} shows how the ground state energy $E(n)$ and the entanglement entropy $S(n)$ vary with the restarted Lanczos number $n$ for the Heisenberg model, obtained with MPS. The entanglement entropy grows very rapidly with $n$ in the first tens of iterations and converges to a constant value in the limit $n\rightarrow\infty$. The ground-state energy is strongly anti-correlated with the entanglement entropy, dropping quickly with $n$. For example, the relative error in the ground-state energy for $L=20$ is reduced by nearly two orders of magnitude for $D=8$ and three orders of magnitude for $D=12$ at $n=300$. The ground-state energy keeps descending with $n$, but with a smaller slope when the entanglement entropy becomes saturated.

By directly comparing the entanglement entropy of the initial MPS, $\vert \Psi_1\rangle$, to the converged LTNS, shown in Fig. \ref{fig:Entropy1D}(a), we find that the latter contains much more entanglement than the former. In particular, the entanglement entropy of the LTNS with $D=8$ already reaches the values calculated with DMRG by keeping as many as 1024 states, which can be regarded as quasi-exact. Furthermore, the entanglement entropy of the LTNS varies logarithmically with the system size, indicating that the LTNS can describe correctly the scaling behavior of a critical system, even though each individual MPS is bounded by the entanglement area law.

\begin{figure}[tbp]
\includegraphics[angle=0,scale=0.25]{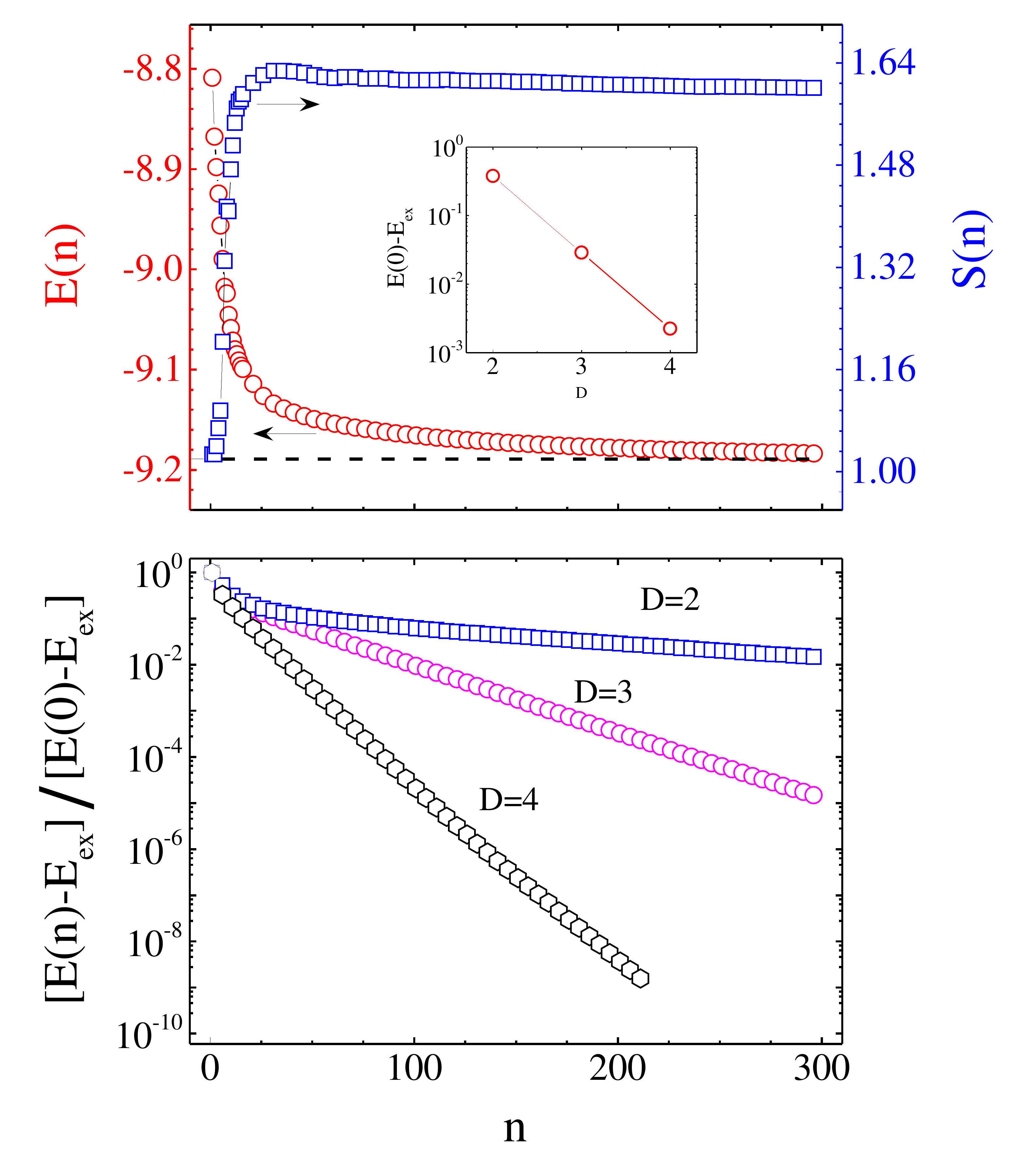}
\caption{Ground-state energy and entanglement entropy for the antiferromagnetic Heisenberg model on the 4$\times$4 square lattice. Upper panel: $E(n)$ and $S(n)$ versus $n$ for $D=2$. The dashed line is the exact ground-state energy, $E_{ex} = -9.1892070651930$. The inset shows the difference $E(0)-E_{ex}$ as a function of $D$. Lower panel: Relative change of the ground state energy, $\left[ E(n) - E_{ex} \right] / \left[ E(0) - E_{ex} \right]$, shown as a function of $n$.}
\label{fig:2D}
\end{figure}

Fig. \ref{fig:Entropy1D}(a) shows the entanglement entropy as a function of the lattice size for the Heisenberg spin chain. For all the lattice sizes we have studied, we find that the converged entanglement entropy agrees with the equation predicted by conformal field theory \cite{2009Calabrese}
\begin{equation}
S = \frac{c}{6} \ln L + b,
\end{equation}
where $b$ is a non-universal constant and $c$ is the central charge. In our calculation, both $c$ and $b$ are weakly dependent on the bond dimension. For $D = 8$ and $12$, $c$ is found to be approximately 1.014, close to the exact result, $c=1$.

\begin{figure}[tbp]
\includegraphics[angle=0,scale=0.23]{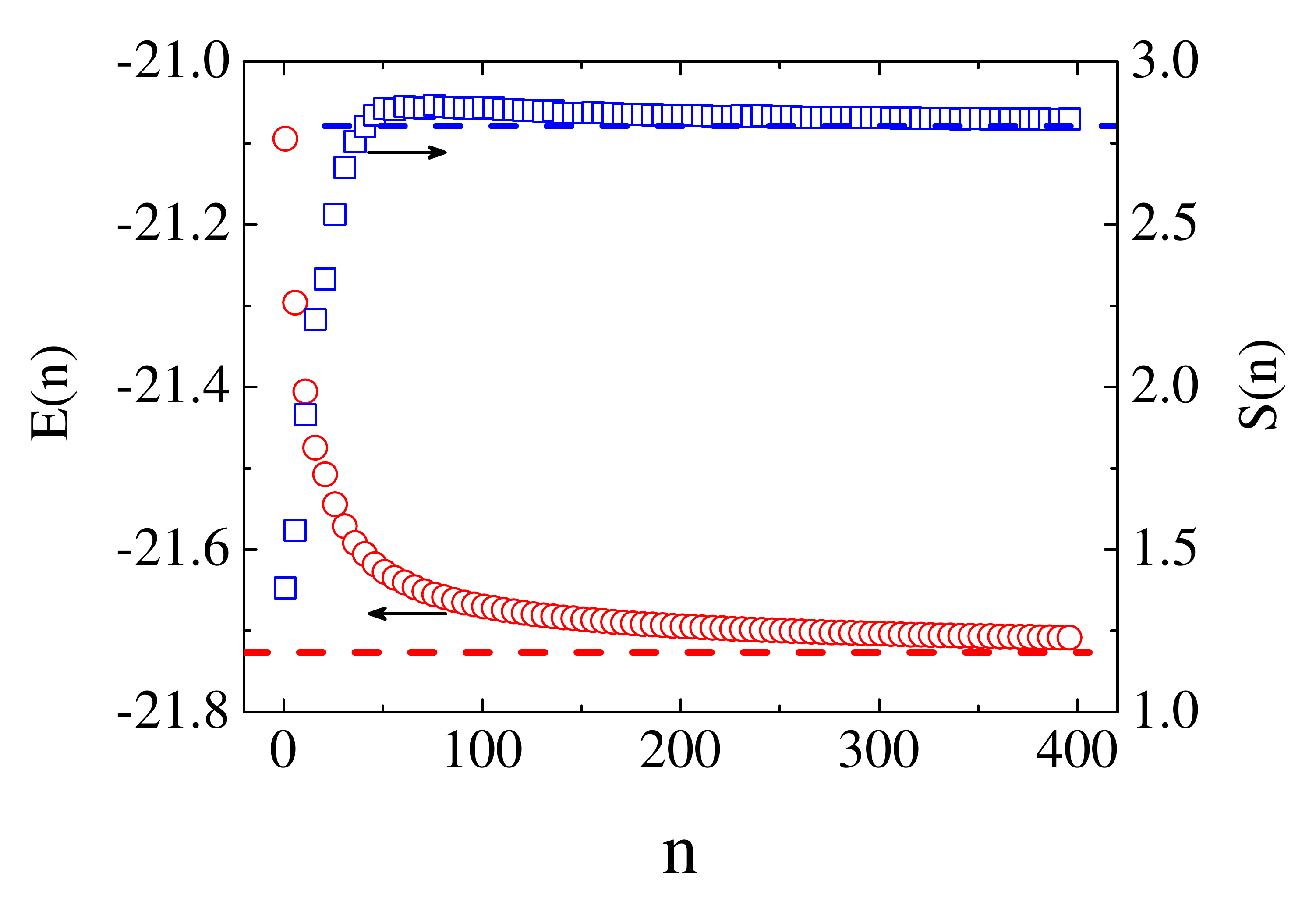}
  \caption{Ground-state energy and entanglement entropy for the Heisenberg model with $D=2$ on the $6\times 6$ lattice. The dashed lines are the corresponding results obtained using DMRG by keeping 4096 states ($E = -21.7267859$
  and $S=2.802147$). }
\label{fig:6by6}
\end{figure}

The entanglement entropy of the LTNS, as revealed by Fig. \ref{fig:Energy1D}(a), is contributed primarily by the first few tens of TNS generated by the Lanczos iterations. To reproduce the logarithmic $L$-dependence, the number of these TNS that have the most significant contribution to the entanglement entropy should increase with the lattice size. This is indeed what we find. Figure \ref{fig:Entropy1D}(b) shows how the entanglement entropy, $S(n)$, approaches its converged value, $S(\infty)$, with $n$ for several different values of $L$. Clearly, the number of TNS that have the most significant contributions to the entanglement entropy increases with $L$. If $n_0$ is the number of TNS whose contribution to the entanglement entropy is within 1\% of the converged value, i.e. $1 - S(n_0)/S(\infty) \le 1\%$, we find that $n_0$ varies almost linearly with the system size (inset, Fig. \ref{fig:Entropy1D}(b)).

In two dimensions, the improvement of the TNS-Lanczos technique over the conventional tensor-network algorithm is even more pronounced. Figure \ref{fig:2D} shows how the ground-state energy and the entanglement entropy vary with the Lanczos number for the isotropic Heisenberg model on the $4\times 4$ square lattice  with open boundary conditions obtained with PEPS.  Similar to  one dimension, the ground-state energy drops very rapidly in the first tens of iterations while the entanglement entropy grows most rapidly.
As the ground state of the two-dimensional Heisenberg model satisfies the entanglement area law, we find that the ground-state energy $E(n)$ converges even faster than its one-dimensional counterpart.  For example, with just 200 Lanczos iterations, the accuracy of the ground-state energy is improved by 9 orders of magnitude for $D=4$. Moreover, we find that the ground state energy converges exponentially to the exact result, $E_{ex}$, in the large $n$ limit
\begin{equation}
  E(n) = E_{ex} + a e^{-\mu n} ,
  \label{Eq:exponential}
\end{equation}
where $a$ and $\mu$ are two parameters depending on both the bond dimension and the system size. $\mu$ measures the rate of convergence. Given $L$, $\mu$ increases rapidly with increasing $D$.

\begin{figure}[tbp]
\includegraphics[angle=0,scale=0.23]{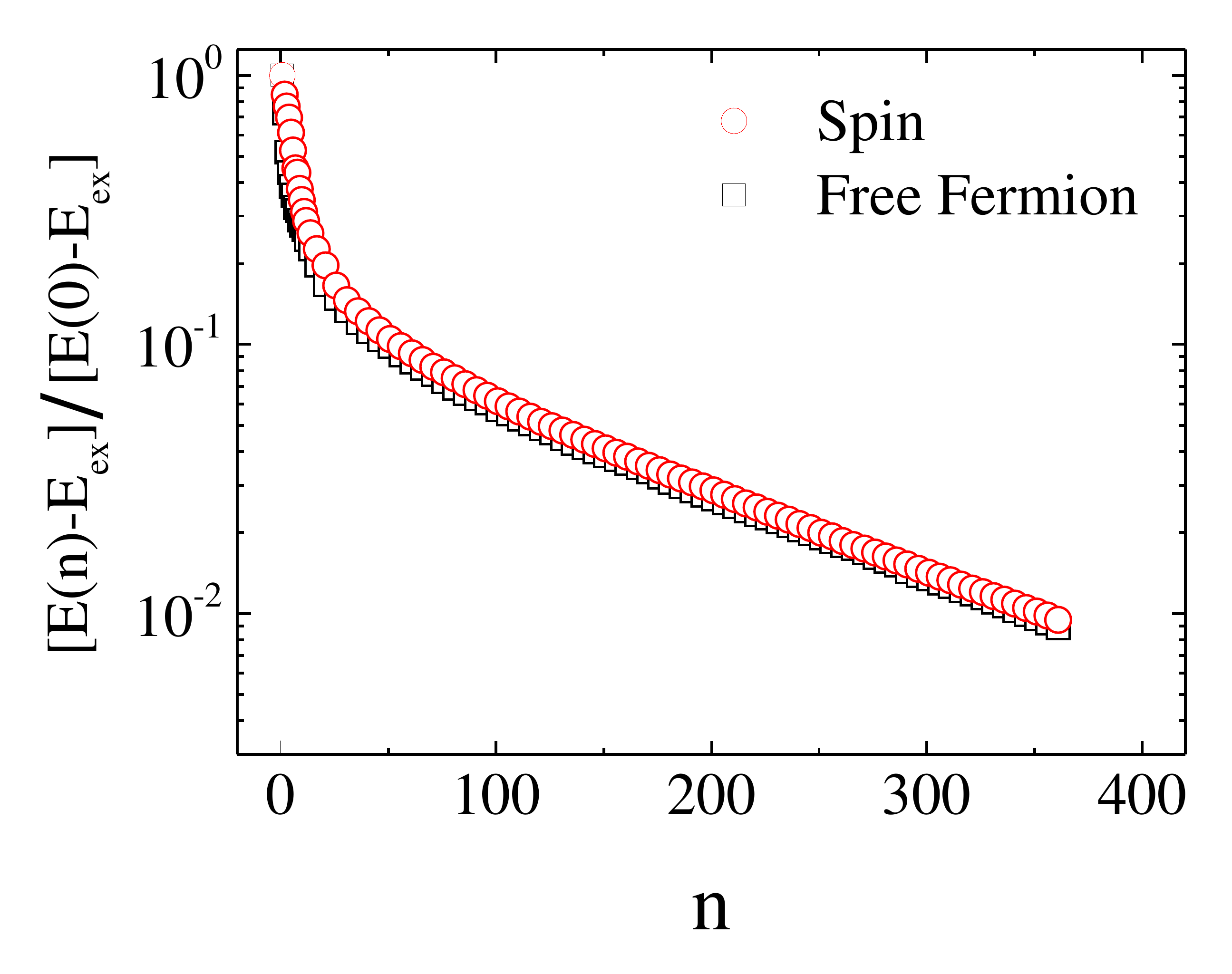}
  \caption{$\left[ E(n)-E_{ex}\right] / \left[ E(0) -E_{ex} \right]$ versus $n$ for the $S=1/2$ free-fermion model on the $4\times 4$ lattice with free boundary conditions with $D=2$, $E(0) - E_{ex}=2.805338\times 10^{-4}$.  The corresponding result for the Heisenberg model is shown for comparison.  }
\label{fig:fermion}
\end{figure}

The method works similarly in larger lattice systems. Fig. \ref{fig:6by6} shows the ground-state energy and entanglement entropy as functions of $n$ for the Heisenberg model on the $6\times 6$ lattice. By comparison, we find that the converged entanglement entropy of the LTNS is larger than the corresponding DMRG result obtained by keeping 4096 states. This implies  that the LTNS we obtain with $D=2$ already contains more information on the entanglement structure of the ground state, and hence is more accurate in describing the correlation functions, than the DMRG wavefunction with $D=4096$.

We have also tested the method for fermionic systems. As revealed by
Fig. \ref{fig:fermion}, the normalized difference in the ground state energy, $\left[ E(n)-E_{ex}\right] / \left[ E(0) -E_{ex} \right]$, for the non-interacting spin-1/2 fermion model drops as fast as in the Heisenberg model on the $4\times 4$ lattice with $D=2$, indicating that the TNS-Lanczos approach works equally well with fermions.

{\it Summary:}
The generalized Lanczos method we propose provides a powerful numerical tool to solve quantum lattice models using TNS. It improves significantly the existing tensor-network algorithm and allows the ground state to be calculated accurately using TNS with small bond dimensions. The ground state wavefunction obtained with this method contains more entanglement than a single TNS. It can describe correctly the logarithmic correction to the area law of entanglement entropy in a critical system without invoking a mulitscale entangled TNS, such as MERA.

The TNS-Lanczos approach can be applied to the MPS, PEPS, MERA or any other kind of TNS with any kind of boundary conditions. It can be extended to calculate the second or even higher excitation states and the energy gap by targeting two or more basis states at each Lanczos iteration. This can be regarded as a generalization of the block Lanczos method. It can be extended to a non-Hamiltonian system to compute, for example, thermodynamic quantities using quantum transfer matrices \cite{1996Bursill,1997Wang}. It can be also extended to compute dynamic quantities, such as the Green function of electrons, using Chebyshev expansion \cite{2011Holzner}. Other kind of Krylov subspace methods similar to the Lanczos method, for example, the Arnoldi and the conjugate-gradient methods, can be also used to generate the Krylov basis states.

This work was supported by the National Natural Science Foundation of China (Grants No. 11190024 and No. 11474331).


\end{document}